\documentclass[sigconf]{acmart}
\usepackage{chngcntr}
\usepackage{graphicx}
\usepackage{amsmath}
\usepackage{subcaption}
\usepackage{amsfonts}
\usepackage{algorithmic}
\usepackage{soul}
\usepackage{url}
\usepackage{balance}
\usepackage{multirow, multicol, makecell}
\usepackage{url} 
\usepackage{xspace}
\usepackage{framed}

\usepackage{enumitem} 
\usepackage[breakable,skins]{tcolorbox}

\newcommand{\ignore}[1]{}

\title{Real-Time Toxicity Filtering for Open-Source Code Reviews}

\author{
Md Awsaf Alam Anindya$^\clubsuit$,
Showvik Biswas$^\clubsuit$,
Anindya Iqbal$^\clubsuit$,
Jaydeb Sarker$^\heartsuit$,
Amiangshu Bosu$^\spadesuit$
}

\affiliation{%
  \institution{$^\clubsuit$ Bangladesh University of Engineering \& Technology \quad
               $^\heartsuit$ University of Nebraska Omaha\quad
               $^\spadesuit$ Wayne State University}
  \country{}
}

\email{
1505114.maaa@ugrad.cse.buet.ac.bd, 1805068@ugrad.cse.buet.ac.bd}
 
\email{anindya@cse.buet.ac.bd,jsarker@unomaha.edu, amiangshu.bosu@wayne.edu
}

\begin{abstract}

\end{abstract}



\copyrightyear{2026}
\acmYear{2026}
\setcopyright{cc}
\setcctype{by}
\acmConference[FSE Companion '26]{34th ACM Joint European Software Engineering Conference and Symposium on the Foundations of Software Engineering}{July 05--09, 2026}{Montreal, QC, Canada}
\acmBooktitle{34th ACM Joint European Software Engineering Conference and Symposium on the Foundations of Software Engineering (FSE Companion '26), July 05--09, 2026, Montreal, QC, Canada}
\acmDOI{10.1145/3803437.3807395}
\acmISBN{979-8-4007-2636-1/2026/07}

\begin{document}

\begin{abstract}

Toxic interactions in open-source software development harm community collaboration. To combat this, we propose \textit{ToxiShield}, a real-time browser extension that identifies and detoxifies toxic code reviews. The framework comprises three modules: toxicity identification, reasoned multiclass classification, and code review detoxification. Our fine-tuned BERT-based binary classifier achieved a 97\% F1-score on 38,761 code review texts. For multiclass classification, Claude 3.5 Sonnet with prompt engineering achieved a 39\% MCC and 42\% F1 on 1,200 samples. Finally, our fine-tuned Llama 3.2 detoxification model reached 95.27\% style transfer accuracy, 97.03\% fluency, 67.07\% content preservation, and an 84\% J-score. Validation with 10 software developers suggests \textit{ToxiShield} effectively fosters a more inclusive open-source environment.
\end{abstract}

\maketitle


\section{Introduction}


Modern software development, particularly within Open Source Software (OSS), relies heavily on collaboration. Platforms such as GitHub facilitate this work by providing features like version control and code review. However, research indicates that such collaboration can often breach professional boundaries and deteriorate into toxic interactions~\cite{sarker2023automated, miller2022did}. The consequences include hindering productivity, creating barriers for newcomers~\cite{miller2022did}, fostering conflicts, and ultimately reducing the quality of the software itself~\cite{sarker2025landscape}.

\begin{figure}
	\centering  
\includegraphics[width=0.9\linewidth, trim=0 0 0 0, clip] {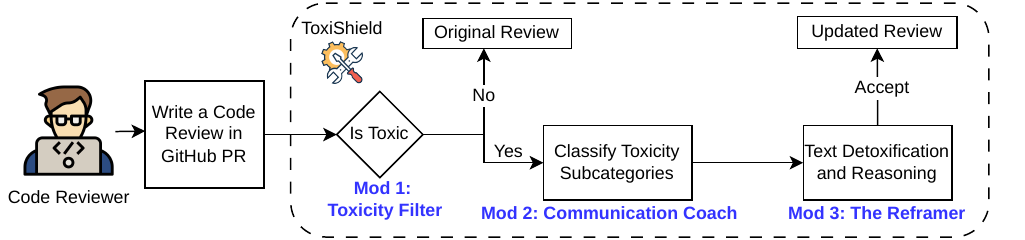}
	\caption{Motivational Workflow of ToxiShield}
	\label{fig:method_flow}	
    \vspace{-12pt}
\end{figure}

Existing approaches to managing OSS toxicity rely on post hoc text classification, often intervening only after detrimental effects have already occurred~\cite{rahmandwhp}. To address this, we aim to develop a tool that not only flags toxic code review comments but also provides actionable feedback and suggests civil alternatives. While recent work by Rahman \textit{et al.} uses a customized T5 model to rephrase uncivil comments~\cite{rahmandwhp}, it lacks explainability and real-time integration. Therefore, we propose \textbf{ToxiShield}~\cite{anindya2026toxishield}, a framework that proactively detects and mitigates toxicity in SE communication in real time through a three-stage process (Figure~\ref{fig:method_flow}). Our evaluations are available in the replication package~\cite{ToxiShield}.


\section{Methodology}

ToxiShield employs a three-module approach to mitigate toxicity in SE communication by (1) detecting toxicity using a binary classifier, (2) sub-categorizing a toxic comment into one or more predefined classes of toxicity, and (3) reframing a toxic comment into a civil, semantically equivalent alternative, additionally providing a detailed explanation of why the comment was flagged as toxic. We detail each of the modules below.

\subsection{Module 1: Toxicity Filter}

We employed Masked Language Models (MLMs) and Large Language Models (LLMs) for classification. Because existing models often struggle with nuanced language like sarcasm in code reviews~\cite{sarker2023automated}, we fine-tuned MLMs on a newly curated dataset. Drawing from Sarker \textit{et al.}'s~\cite{sarker2025landscape} study of 101 million OSS comments, we filtered 15 million pull-request comments to annotate 10,120 toxic and 28,641 non-toxic instances. We fine-tuned \textit{BERT-base-uncased} and \textit{Xtreme-DistilBERT} using an 8:1:1 split (train: validation: test) and 10-fold stratified cross-validation to select the optimal checkpoint. Also, we evaluated a zero-shot prompted GPT-4o on the test set. 

\textbf{Results:} The fine-tuned \textit{BERT-base-uncased} model yielded the best results, achieving 0.98 precision, 0.96 recall, and a 0.97 toxic-class F1-score. Error analysis highlighted misinterpretations of domain-specific terms as profanity, tokenization issues, and failures in detecting passive-aggressiveness and mockery.

\subsection{Module 2: Communication Coach}

Once a comment is classified as toxic by the Toxicity Filter, it is passed to the communication coach, which categorizes the text into one or more of 11 toxicity sub-categories and provides targeted explanatory feedback. For this task, we adopted the taxonomy of toxicity sub-categories proposed by Sarker \textit{et al.} Given the scarcity of fine-grained annotated data, we combined their dataset of 600 manually annotated comments (which included 532 toxic samples) with 600 non-toxic comments from our baseline dataset, yielding a multi-label, multi-class dataset of 1,200 code review comments. Since data scarcity prevented fine-tuning of traditional language models, we leveraged LLMs for their strong performance on prompt-based multi-class tasks~\cite{dev-esem-2025}. We incorporated \textit{in-context learning} via prompt engineering, a technique that has shown efficacy in both general and toxic text classification. Our iterative prompting strategy began with a base prompt containing five elements: (i) role clarification, (ii) task assignment, (iii) category definitions and examples, (iv) explicit operational guidelines, and (v) a strict XML output specification. We refined these prompts across five stages, progressively conditioning the model to adopt the persona of an experienced OSS maintainer. Each stage was rigorously evaluated against a validation set using the Matthews Correlation Coefficient (MCC). We performed this prompt optimization using GPT-4o. Subsequently, we ran Claude 3.5 Sonnet and Llama 3.3 70B using the optimized prompt, employing deterministic decoding (temperature = 0) to ensure reproducibility.

\textbf{Results:} We evaluated model performance using Macro-F1 and Macro-MCC metrics. Claude 3.5 Sonnet achieved the highest performance, achieving a Macro-F1 score of 0.42 and a Macro-MCC of 0.39. An analysis of the classification results indicated that while the models accurately detected explicit sub-categories, such as Profanity, they struggled with more ambiguous cases. For example, Object-Directed Toxicity was frequently misclassified as Non-Toxic, Profanity, Insult, or Trolling. To complement our quantitative metrics and validate the classifications, two authors independently reviewed a subset of 100 instances to calculate inter-rater reliability. This reliability, measured using Cohen's $\kappa$, was calculated at 0.66, indicating substantial agreement between the reviewers.

\section{Module 3: The Reframer}

We developed this `Reframer Module' as a specialized real-time detoxification mechanism that automatically rephrases hostile communication in OSS environments into constructive alternatives while preserving its original technical substance. We formulate the detoxification of code reviews as a Text Style Transfer (TST) task, aiming to shift the stylistic tone from toxic to neutral or professional while rigorously preserving the underlying technical semantics. This requires a parallel corpus of toxic comments paired with faithful, non-toxic rewrites~\cite {fu2018style}. Due to the lack of such a corpus for the SE domain, we synthesised a parallel dataset. We used the 10,120 toxic comments identified in the dataset from Module 1 as the source data. To generate their non-toxic counterparts, we leveraged multiple state-of-the-art (SoTA) LLMs by prompting them to rewrite toxic comments into respectful alternatives, strictly adhering to the original intent. We iteratively refined the prompt templates using Chain-of-Thought (CoT) reasoning against our primary evaluation metric (J-score) until achieving stable performance. \textit{GPT-4o-05-13} achieved the highest J-score of 88.14\%, indicating the best overall balance across metrics. Consequently, we selected the dataset generated by this model as the final parallel corpus for training the student models. \textit{Qwen 2.5 Instruct 7B} followed closely with a J-score of 83.73\%, while the proprietary \textit{GPT-4o mini} achieved a comparable 83.57\%.

\textbf{Results:} We evaluated the performance of our distilled student models using the same metrics applied in the teacher phase, ensuring direct comparability. \textit{Llama 3.2 3B} emerged as the top-performing model, achieving the highest aggregate J-score of 84\%. This model effectively balanced strong style transfer accuracy (95.27\%) and fluency (97.03\%) with respectable content preservation (67.07\%). A manual error analysis of our best-performing model revealed that it frequently struggled to detoxify code reviews involving complex technical context. For example, the comment, \textit{``I'd change it to `is\_disgusting\_for', as the current name implies that the item itself is disgusted by someone picking it up,''} was incorrectly identified as toxic simply because the suggested variable name contained the word \textit{disgusting}.


\section{User Evaluation} 
We integrated all three modules into a browser extension to function as a real-time detoxification tool. Furthermore, we evaluated the tool with 10 professional developers over a two-week period using the Technology Acceptance Model (TAM). The TAM analysis demonstrated that ToxiShield was generally well-received by professional developers, showing particular strengths in ease of use and overall satisfaction. While the tool’s utility was widely acknowledged, enhancing task efficiency and error management mechanisms will be critical for driving broader adoption~\cite{ToxiShield}.

\section{Conclusion}
This is the first study to provide a real-time detoxification tool for developers, providing an explanation. We provide the dataset, code, and tool publicly available for use~\cite{ToxiShield} and DOI: \url{https://doi.org/10.5281/zenodo.19490337}. We hope that \textit{ToxiShield} will help foster inclusiveness in the open-source community by detoxifying toxic communication. 

\bibliographystyle{ACM-Reference-Format}  
\balance
\bibliography{references}

\end{document}